\documentstyle[12pt,epsf]{article}
\def\selecmp{\tilde e^{\mp}}

\def\selecR{\tilde e_{\rm R}}
\def\selecL{\tilde e_{\rm L}}

\def\lsp{{\tilde\chi^0}_1}
\def\cinolm{\tilde{\chi_{\rm 1}}^-}
\def\cinolp{\tilde{\chi_{\rm 1}}^+}
\def\photino{\tilde\gamma}

\begin{document}
\pagestyle{plain}
\renewcommand{\thefootnote}{*)}

\title{\bf
\begin{quote}
 \raggedleft TMCP--95--3\\
 March 1995\\ ~
\end{quote}
The GRACE system for SUSY processes}
\font\fonts=cmbx12
\author{~\\
\bf Masato J{\fonts IMBO}\\
\it Computer Science Laboratory, Tokyo Management College,\\
\it Ichikawa, Chiba 272, Japan\\
\it (e-mail: jimbo@kekvax.kek.jp)\\
\and \bf M{\fonts INAMI}-T{\fonts ATEYA} collaboration\\
}\date{ }
\maketitle
\begin{abstract}
    We introduce a new method to treat Majorana fermions
on the GRACE system which has been developed for the
automatic computation of the matrix elements for the
processes of the standard model.  In the GRACE system, we
already have such particles as Dirac fermions, gauge
bosons and scalar bosons within the standard model.  On
the other hand, in the SUSY models there are Majorana
fermions.  In the first instance, we have constructed a
system for the automatic computation of cross-sections for
the processes of the SUSY QED.  The system has also been
applied to another model including Majorana fermions, the
minimal SUSY standard model (MSSM), by the re-definition
of the model file.
\end{abstract}
\renewcommand{\thefootnote}{\sharp\arabic{footnote}}
\renewcommand{\theequation}{\arabic{section}.\arabic{equation}}
\setcounter{footnote}{0}

\section{Introduction}

    It has been a promising hypothesis that there exists a symmetry called
supersymmetry (SUSY) between bosons and fermions at the unification-energy
scale.  It, however, is a broken symmetry at the electroweak-energy scale.
The relic of SUSY is expected to remain as a rich spectrum of SUSY particles,
partners of usual matter fermions, gauge bosons and Higgs scalars, named
sfermions, gauginos and higgsinos, respectively~\cite{theor}.

    The quest of these SUSY particles has already been one of the most
important
pursuits to the present high-energy physics~\cite{exp}.  Although such
particles have not yet been discovered, masses of them are expected to be
$O(10^2)$ GeV~\cite{tub}.  In order to obtain signatures of the SUSY-particle
production, electron-positron colliding experiments are preferable because the
electroweak interactions are clean and well-known.  Thus we hope SUSY particles
will be found out at future TeV-region (sub-TeV region) $e^-e^+$-colliders such
as CLIC, NLC and JLC~\cite{col}.

    For the simulations of the experiments, we have to calculate the
cross-sections for the processes with the final 3-body or more.  We have
already known within the standard model that the calculation of the helicity
amplitudes is more advantageous to such a case than that of the traces for the
gamma matrices with REDUCE~\cite{tks,wmh}.  The program package
CHANEL~\cite{chan} is one of the utilities for the numerical calculation of the
helicity amplitudes.

    It, however, is also hard work to construct a program with many subroutine
calls of CHANEL by hand.  Thus we need a more convenient way to carry out such
a work.  The GRACE system~\cite{pre-grc}, which automatically generates the
source code for CHANEL, is one of the solutions.  The system also includes the
interface and the library of CHANEL, and the multi-dimensional integration
and event-generation package BASES/SPRING v5.1~\cite{bas}.

\begin{figure}[htp]
\setlength{\unitlength}{1mm}
\begin{picture}(150,180)

\put(80,170){\framebox(50,8)[c]{\hbox{Theory(Lagrangian)}}}
\put(10,170){\framebox(50,8)[c]{\hbox{User input}}}
\put(10,155){\framebox(50,15)[c]{\hbox{Process(particle, order)}}}
\put(80,155){\framebox(50,8)[c]{\hbox{Particles and interactions}}}

\put(105,170){\vector(0,-1){7}}
\put(105,155){\vector(0,-1){5}}
\put(35,155){\vector(0,-1){5}}
\put(20,155){\vector(0,-1){40}}

\put(31,141){\framebox(78,8)[c]{\  }}
\put(30,140){\framebox(80,10)[c]{\hbox{Diagram generator}}}

\put(70,140){\vector(0,-1){7}}
\put(70,125){\vector(0,-1){25}}

  \put(50,125){\framebox(40,8)[c]{\hbox{Diagram description}}}
  \put(101,126){\framebox(38,8)[c]{ }}
  \put(100,125){\framebox(40,10)[c]{\hbox{Drawer}}}
  \put(100,105){\dashbox(40,10)[c]{\hbox{Feynman diagrams}}}
    \put(90,130){\vector(1,0){10}}
    \put(105,125){\vector(0,-1){10}}

  \put(5,105){\framebox(40,10)[c]{}}
   \put(6,108){\hbox{Kinematics database}}
  \put(23,105){\vector(0,-1){26}}

\put(31,91){\framebox(78,8)[c]{\  }}
\put(30,90){\framebox(80,10)[c]{\hbox{Matrix element generator}}}

  \put(15,65){\framebox(120,20)[c]{\  }}
    \put(20,67){\framebox(25,12)[c]{\  }}
    \put(54,67){\framebox(27,12)[c]{\  }}
    \put(90,67){\framebox(25,12)[c]{\  }}
    \put(28,80){\hbox{Function for matrix element}}
    \put(80,80){\hbox{\footnotesize (FORTRAN)}}
    \put(22,69){\hbox{code}}
    \put(22,73){\hbox{kinematics}}
    \put(57,69){\hbox{code}}
    \put(57,73){\hbox{generated}}
    \put(92,69){\hbox{library}}
    \put(92,73){\hbox{CHANEL}}
    \put(45,73){\line(1,0){9}}
    \put(90,73){\line(-1,0){9}}

\put(70,90){\vector(0,-1){11}}
\put(70,65){\vector(0,-1){5}}

\put(41,51){\framebox(78,8)[c]{\  }}
\put(40,50){\framebox(80,10)[c]{\hbox{BASES(Monte-Carlo integral)}}}

  \put(5,50){\framebox(23,10){\  }}
    \put(6,51){\hbox{information}}
    \put(6,55){\hbox{convergence}}
  \put(40,55){\vector(-1,0){12}}

  \put(5,28){\dashbox(45,15){\  }}
    \put(15,32){\hbox{Distributions}}
    \put(15,37){\hbox{Cross section}}
  \put(45,50){\vector(0,-1){7}}

  \put(55,50){\vector(0,-1){26}}
  \put(20,9){\framebox(40,15){\  }}
    \put(22,13){\hbox{Parameters}}
    \put(22,18){\hbox{Distribution}}
  \put(60,22){\vector(1,0){10}}

\put(71,21){\framebox(78,8)[c]{\  }}
\put(70,20){\framebox(80,10)[c]{\hbox{SPRING(event generator)}}}

\put(130,65){\vector(0,-1){35}}
\put(110,20){\vector(0,-1){5}}

  \put(85,5){\dashbox(50,10){\  }}
    \put(95,6){\hbox{generated events}}
    \put(95,10){\hbox{Specified number of}}
\end{picture}
  \hspace*{2cm} Fig.~1. GRACE system flow
\end{figure}

    In the SUSY models, there exist Majorana fermions as the neutral gauginos
and higgsinos, which become the mixed states called neutralinos.  Since
anti-particles of Majorana fermions are themselves, there exists so-called
`Majorana-flip', the transition between particle and anti-particle.  This has
been the most important problem which we should solve when we realize the
automatic system for computation of the SUSY processes.

    In a recent work~\cite{jt,jtkk}, we developed an algorithm to treat
Majorana fermions in CHANEL.  In the standard model, we already have such
particles as Dirac fermions, gauge bosons and scalar bosons in the GRACE
system.  Thus we can construct an automatic system for the computation of the
SUSY processes by the algorithm above in the GRACE system.  In this work, we
present the check list of the system at this time, and one of the results.

\section{Majorana fermions into new GRACE}

    In Fig.~1, we present the system flow of GRACE~\cite{gm}.  The GRACE system
has become more flexible for the extension in the new version called
`{\bf grc}'~\cite{grc-pp}, which includes a new graph-generation package.  With
this package, every graphs can be generated based on a user-defined model.  It
is necessary for us to make the interface and the library of CHANEL and the
model file for including the SUSY particles.

   The method of computation in the program package CHANEL is as follows:
\begin{enumerate}
  \item To divide a helicity amplitude into vertex amplitudes.
  \item To calculate each vertex amplitude numerically as a complex number.
  \item To reconstruct of them with the polarization sum, and calculate
  the helicity amplitudes numerically.
\end{enumerate}
The merit of this method is that the extension of the package is easy,
and that each vertex can be defined only by the type of concerned particles.

    Here we propose an algorithm~\cite{jt,jtkk} for the implementation of the
embedding Majorana fermions in CHANEL as follows:
\begin{itemize}
  \item \underline{\bf policy}
  \begin{enumerate}
    \item To calculate a helicity amplitude numerically.
    \item To replace each propagator by wave functions or polarization vectors,
    and calculate vertex amplitudes.
    \item \underline{\bf Not to} move charge-conjugation matrices.
  \end{enumerate}
    \item \underline{\bf method}
  \begin{enumerate}
    \item To choose a direction on a fermion line.
    \item To put wave functions, vertices and propagators along the direction
    in such a way:
  \begin{itemize}
    \item[~i)] To take the transpose for the reverse direction of
    fermions
    \item[ii)] To use the propagator with the charge-conjugation
    matrix for\\
    the Majorana-flipped one.
  \end{itemize}
  \end{enumerate}
\end{itemize}
As a result, the kinds of the Dirac-Majorana-scalar vertices are limited to
four types:
\begin{itemize}
\begin{itemize}
  \item[(1)] $\overline{U} \Gamma U$
  \item[(2)] $U^{\rm T} \Gamma \overline{U}~^{\rm T}$
  \item[(3)] $\overline{U} C^{\rm T} \Gamma^{\rm T} \overline{U}~^{\rm T}$
  \item[(4)] $U^{\rm T} \Gamma^{\rm T} C^{-1} U$
\end{itemize}
\end{itemize}
where $U$'s denote wave functions symbolically without their indices, and $C$
is the charge-conjugation matrix. The symbol $\Gamma$ stands for the scalar
vertex such as
\[ \Gamma = A_{\rm L}\cdot{{1 - \gamma}\over{2}} +
A_{\rm R}\cdot{{1 + \gamma}\over{2}} ~~. \]
The vertices (2)$\sim$(3) are related to the vertex (1) which is the same
as the Dirac-Dirac-scalar vertex in the subroutine of CHANEL.  Thus we can
build three new subroutines for the added vertices.  We have performed the
installation of the subroutines above with the interface on the new GRACE
system.

\section{Numerical results}

    At the start for the check of our system, we have written the model file
of the SUSY QED.  In this case, there is only one Majorana fermion, photino.
Next we have extended the model file and the definition file of couplings for
the MSSM.  The tests have been performed by the exact calculations with the two
methods, our system and REDUCE.  In Table I, the tested processes are shown as
a list.  The references in the table (without \cite{jtkk}) are not the results
of the tests, but for help.

\begin{table}[hbt]
\begin{center}
  \begin{tabular}{llclcc}  \hline
  Process &  & Number of diagrams & Comment & Check & Reference \\
  \hline\hline
  {\bf SUSY} & {\bf QED} & & & & \\ \hline
$e^- e^- \rightarrow$ & $\selecR^- \selecR^-$ & 2 & Majorana-flip & OK & \\
        & $\selecL^- \selecL^-$ & 2 & in internal lines & OK & \cite{jtkk} \\
                    & $\selecR^- \selecL^-$ & 2 & & OK & \\  \hline
$e^- e^+ \rightarrow$ & $\selecR^- \selecR^+$ & 2 & Including pair & OK &
		    \cite{mj}\\
        & $\selecL^- \selecL^+$ & 2 & annihilation & OK & \cite{mj} \\ \hline
$e^- e^+ \rightarrow$ & $\selecR^- \selecL^+$ & 1 & Values are & OK &
		    \cite{mj} \\
        & $\selecR^+ \selecL^-$ & 1 & equal & OK & \cite{mj} \\  \hline
$e^- e^+ \rightarrow$ & $\photino \photino$ & 4 & F-B symmetric & OK &
	\cite{jtkk} \\  \hline
$e^- e^+ \rightarrow$ & $\photino \photino \gamma~$ & 12 & Final 3-body
   & OK & \cite{tk} \\ \hline
$e^- e^+ \rightarrow$ & $\selecR^- \photino e^+$ & 12 & Including every
   & OK & \cite{spu} \\
   & & & elements for tests & & \cite{epa} \\ \hline\hline
  {\bf MSSM} & & & & \\ \hline
$e^- e^- \rightarrow$ & $\selecL^- \selecL^-$ & 8 & 4 Majorana fermions & OK &
   \\ \hline
$e^- e^+ \rightarrow$ & $\cinolm \cinolp$ & 3 & & OK &\\  \hline
  \end{tabular}
\end{center}
  \hspace*{4cm} Table~I. The list of the tested processes.
\end{table}%

    Here we present the results for the single-selectron production within the
SUSY QED process at the energy $\sqrt{s} = 190$ GeV.  The masses of the
concerned particles are $M_{\photino} = 50$ GeV, $M_{\selecR} = 100$ GeV and
$M_{\selecL} = 130$ GeV in the calculation.  This is the case that the
pair-production processes occur for both selectrons at the JLC-I energy, but
they do not at the LEP-II energy.  The Feynman diagrams for this process, which
are drawn by the program package `{\bf gracefig}'~\cite{sk} in the new GRACE,
are shown in Fig.~2.

%

    In Fig.~3, we show the angular distribution of the outgoing positron in the
process $e^- e^+ \rightarrow \selecR^- \photino e^+$.  Here we use BASES for
the calculation from the REDUCE output. The result is in beautiful agreement
with the value that is obtained by GRACE at each bin of the histogram.
Since the two diagrams with the one-photon exchange dominate in this case,
there is a steep peak in the direction of the initial positron.  In such a
case, the equivalent-photon approximation (EPA) works well~\cite{epa}.

%

    In Fig.~4, we show the $T_S$ distribution of the selectron in the process
$e^- e^+ \rightarrow \selecR^- \photino e^+$.  The quantity $T_S$ is defined
in Ref.~\cite{epa} as
\[ T_S = {{{P_T}^2 + {M_{\selecR}}^2}\over{{M_{\selecR}}^2}} ~~,  \]
where $P_T$ denotes the transverse momentum of the selectron.  We show also the
$P_T$ distribution of the selectron in the process $e^- e^+ \rightarrow
\selecR^- \photino e^+$ in Fig.~5.  Here we calculate the two dominant diagrams
for comparison with the result from EPA.

%

%

\section{Summary}

    We introduce a new method to treat Majorana fermions on the GRACE system
for the automatic computation of the matrix elements for the processes of the
SUSY models.  In the first instance, we have constructed the system for the
processes of the SUSY QED because we should test our algorithm for the
simplest case.  The numerical results convince us that our algorithm is
correct.

    It is remarkable that our system is also applicable to another model
including Majorana fermions ({\it e.g.} the MSSM) once the definition of the
model file is given.  We have calculated the processes $e^- e^+ \rightarrow
\photino \photino \gamma$ and $e^- e^+ \rightarrow \selecR^- \photino e^+$
within the SUSY QED.  We should calculate the single-photon event from $e^- e^+
\rightarrow \lsp \lsp \gamma$~\cite{tk}, and the resultant single-positron
(electron) event from the single-selectron production $e^- e^+ \rightarrow
\selecmp \lsp e^\pm$~\cite{spu} as soon as possible.  It should be emphasized
that the GRACE system including SUSY particles is the powerful tool for this
purpose.

\section{Acknowledgements}

  This work was supported in part by the Ministry of Education,
  Science and Culture, Japan under Grant-in-Aid for International
  Scientific Research Program No.04044158.  One of the author (M.J.) and Dr.
  H. Tanaka have been also indebted to the above-mentioned Ministry under
  Grant-in-Aid No.06640411.

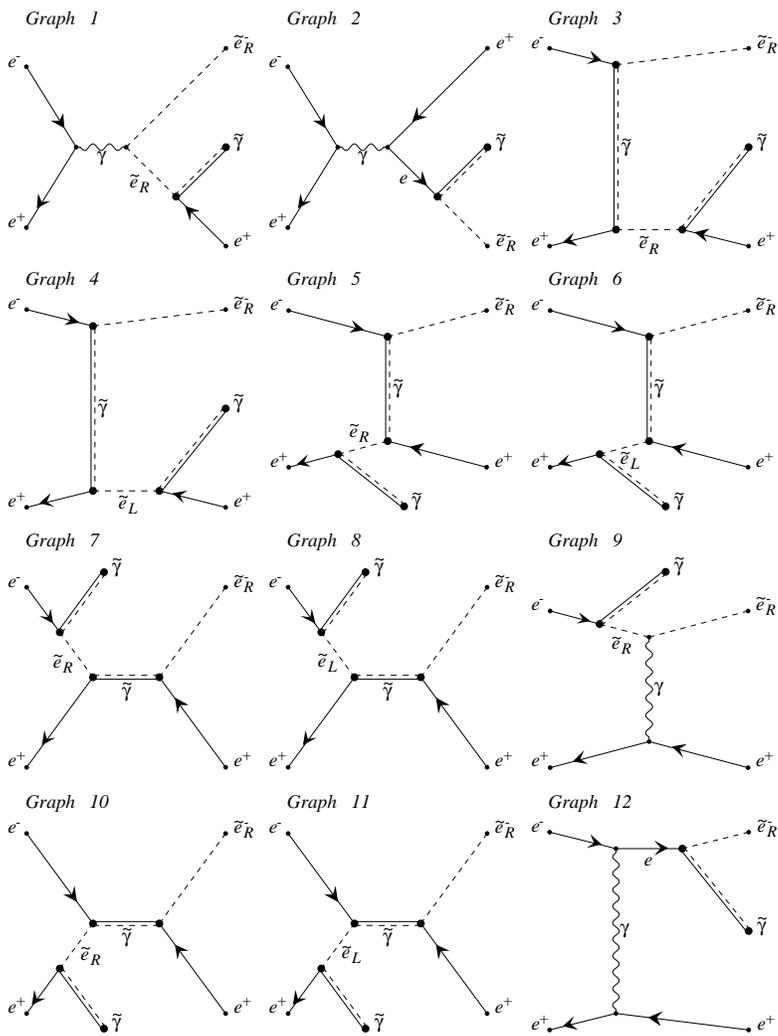
\end{document}